\documentclass[aps,preprintnumbers,amsmath,amssymb,superscriptaddress,floatfix,a4paper,nofootinbib,11pt,floatfix,nofootinbib,onecolumn]{revtex4}
\usepackage{graphicx}
\usepackage{bm}
\usepackage{rotating}
\usepackage{array}
\usepackage{xcolor}
\usepackage{amsmath,amssymb}
\usepackage{mathrsfs}
\usepackage{graphicx}
\usepackage{color}
\usepackage{subfigure}
\usepackage{fancyhdr}
\usepackage{multirow}
\usepackage{float}
\usepackage{epsfig}
\usepackage{amsfonts}
\usepackage{bm}
\usepackage{multirow}
\usepackage{nicefrac}
\usepackage{cancel}
\newcommand{\ba}{\begin{eqnarray}}
\newcommand{\ea}{\end{eqnarray}}

\usepackage{bbm}
\newcommand{\be}{\begin{eqnarray}}
\newcommand{\ee}{\end{eqnarray}}
\usepackage{slashed}

\begin{document}

\title{Scalar Field Dynamics in Non-Minimally Coupled Theories via the Noether Symmetry and the Eisenhart-Duval Lift}

\author{Ahmadfikri Talek}
\email{fikri.talek@gmail.com}
\affiliation{Division of Physical Science, Faculty of Science, Prince of Songkla University, Hatyai 90112, Thailand} 

\author{Narakorn Kaewkhao}
\email{naragorn.k@psu.ac.th}
\affiliation{Division of Physical Science, Faculty of Science, Prince of Songkla University, Hatyai 90112, Thailand} 

\author{Watcharakorn Srikom}
\email{watcharakorn.sri@sru.ac.th}
\affiliation{Department of Innovative Technology for Renewable Energy, Faculty of Science and Technology, Suratthani Rajabhat University, Surat Thani, 84100, Thailand}

\author{Farruh Atamurotov}
\email{atamurotov@yahoo.com}
\affiliation{University of Tashkent for Applied Sciences, Str. Gavhar 1, Tashkent 100149, Uzbekistan}
\affiliation{Urgench State University, Kh. Alimdjan str. 14, Urgench 220100, Uzbekistan}

\author{Phongpichit Channuie} 
\email{phongpichit.ch@mail.wu.ac.th}
\affiliation{College of Graduate Studies, Walailak University, Thasala, Nakhon Si Thammarat, 80160, Thailand}
\affiliation{School of Science, Walailak University, Thasala, \\Nakhon Si Thammarat, 80160, Thailand}

\date{\today}

\begin{abstract}

This study investigates the dynamics of a non-minimally coupled (NMC) scalar field in modified gravity, employing the Noether gauge symmetry (NGS) approach to systematically derive exact cosmological solutions. By formulating a point-like Lagrangian and analyzing the corresponding Euler–Lagrange equations, conserved quantities were identified, reducing the complexity of the dynamical system. Through the application of Noether symmetry principles, the scalar field potential was found to follow a power-law form, explicitly dependent on the coupling parameter $\xi$, influencing the evolution of the universe. The study further explores inflationary dynamics, showing that for specific values of $\xi$, the potential resembles the Higgs-like structure, contributing to a deeper understanding of early cosmic expansion. To enhance the theoretical framework, the Eisenhart lift method was introduced, providing a geometric interpretation of the system by embedding the dynamical variables within an extended field space. This approach established a connection between the kinetic terms and Killing vectors, offering an alternative perspective on the conserved quantities. The study also derived geodesic equations governing the evolution of the system, reinforcing the link between symmetry-based techniques and fundamental cosmological properties.

\end{abstract}

\maketitle


\section{Introduction}

Astrophysical observations—including those from Type Ia Supernovae \cite{SupernovaCosmologyProject:1998vns,SupernovaSearchTeam:1998fmf}, cosmic microwave background (CMB) radiation \cite{BICEP2:2015xme,WMAP:2010qai,WMAP:2012nax,Planck:2015fie,Planck:2015sxf,BICEP2:2014owc,BICEP2:2015nss}, large-scale structure surveys \cite{SDSS:2003eyi}, baryon acoustic oscillations (BAO) \cite{SDSS:2004kqt}, and weak gravitational lensing \cite{SDSS:2005xqv}—consistently support the picture of an accelerating universe. Despite the remarkable success of the Lambda Cold Dark Matter ($\Lambda$CDM) model \cite{Planck:2018nkj}, it faces persistent theoretical challenges, notably the cosmological constant problem \cite{Weinberg:2000yb} and the coincidence problem \cite{Velten:2014nra}. In light of these issues, the phenomenon of late-time cosmic acceleration has gained considerable attention, with dark energy (DE) proposed as a possible explanation within the framework of general relativity. Alternatively, modifying Einstein’s gravitational theory on cosmological scales has also emerged as a compelling route. Several significant works have explored these directions, including \cite{Nojiri:2010wj,Nojiri:2006ri,Capozziello:2011et,Bamba:2015uma} and references therein. Nevertheless, the fundamental nature of the dark energy sector remains largely unknown and represents a central open question in modern cosmology.

Various theoretical approaches have been proposed to address the dark sector, either by modifying the geometric side of the Einstein field equations or altering the stress-energy tensor \cite{Copeland:2006wr}. Among these, $f(R)$ gravity stands out as one of the simplest and most studied extensions of general relativity, where the Lagrangian is generalized to an arbitrary function of the Ricci scalar $R$ \cite{Bergmann:1968ve,Buchdahl:1970ynr}. Comprehensive analyses of $f(R)$ models are available in \cite{Sotiriou:2008rp, DeFelice:2010aj}, alongside investigations of Born–Infeld-inspired gravitational modifications \cite{BeltranJimenez:2017doy}. In \cite{Nojiri:2017ncd,Odintsov:2023weg}, the cosmological implications of such theories—particularly in the contexts of inflation, bouncing universes, and late-time evolution—are systematically explored. In addition to these approaches, theories involving non-minimal derivative couplings (NMDC) to gravity have garnered increasing interest from both theoretical and phenomenological standpoints \cite{Amendola:1993uh, Capozziello:1999xt, Capozziello:1999uwa, Granda:2010hb, Granda:2010ex,Sushkov:2009hk, Saridakis:2010mf, Skugoreva:2013ooa, Sushkov:2012za, Koutsoumbas:2013boa, Gumjudpai:2015vio, Gumjudpai:2016frh}. Such theories have been investigated for their ability to describe the inflationary epoch and its consequences \cite{Granda:2011zk,Germani:2010gm,Sadjadi:2012zp,Tsujikawa:2012mk,Ema:2015oaa,Yang:2015pga,Myung:2015tga, MohseniSadjadi:2013iou}.

The Noether symmetry approach has become a powerful tool in cosmology for identifying viable models and obtaining exact solutions \cite{Capozziello:1994dn}. When the condition $X^{[1]}_{\rm NGS}{\cal L}=0$ holds, a Noether symmetry is said to exist. This condition allows for the determination of conserved quantities and unknown functions within the theory. A wide range of cosmological scenarios has been studied using this approach, including nonlocal $f(T)$ gravity \cite{Channuie:2017txg,Nurbaki:2020dgw,Jamil:2012fs}, viable mimetic $f(R)$ and $f(R,T)$ models \cite{Momeni:2015gka}, $f(R)$ cosmology \cite{Capozziello:2008ch}, cosmological $\alpha$-attractors \cite{Kaewkhao:2017evn}, scalar–tensor theories \cite{Terzis:2014cra}, the generalized Brans–Dicke model \cite{APaliathanasis_2025}, Horndeski gravity \cite{M. Miranda_2024}, and $f(G)$ gravity \cite{Bajardi:2020osh}. Moreover, exact solutions for potentials, scalar fields, and scale factors have been obtained in anisotropic Bianchi cosmologies \cite{Camci:2007zz,Jamil:2012zm,Channuie:2018now}. Applications also include static cylindrically symmetric spacetimes within $f(R)$ gravity using the Noether symmetry technique \cite{Oz:2021nmj}.

A generalization of this technique, known as the Noether Gauge Symmetry (NGS) approach, has also been developed and applied in cosmological contexts \cite{Palia and Tsamparlis2018,Aslam:2013pga,Hussain:2011wa,Kucuakca:2011np}. This extension incorporates the Rund–Trautmann identity, as given in Eq. (\ref{NGS gaugcd}), which accounts for a non-vanishing gauge term $G$ and an undetermined parameter $\tau$. Recent cosmological studies employing the NGS framework include \cite{Paliathanasis_2024b,Laya2023a,Laya2023b,Laya2023c, Nawazish2024}. The authors of this work have previously explored the NGS approach in various gravity theories. For example, Ref. \cite{Kanesom:2021ytb} discussed the NGS framework within the Eddington-inspired Born–Infeld theory. In Ref. \cite{Dolohtahe:2021aah}, a formal treatment of NMDC gravity using the NGS method was presented. That study focused on deriving the point-like Lagrangian for the Einstein–Hilbert action incorporating NMDC in a spatially flat FLRW background. The model included a scalar field and matter content, aiming to quantify the influence of the scalar field’s kinetic term on cosmic evolution.

This paper is organized as follows: We will start by making a short recap of a formal framework of the non-minimally
coupled scalar field to gravity in the Jordan Frame and study the point-like Lagrangian for underlying theory in Section \ref{ch2}. In Section \ref{ch3}, we study a Hessian matrix and quantify the Euler-Lagrange equations. In Section \ref{ch4}, we apply the NGS approach to the point-like descriptions of the non-minimally coupled scalar field to gravity. We examine particularly exact cosmological solutions. In Section \ref{ch5}, we apply the Eisenhart lift methodology to construct the purely kinematic terms, allowing us to establish a one-to-one correspondence with the Killing equations and examine the solutions they provide. This symmetry guides the selection of the potential term and reveals the relationships between the dynamical variables. Finally, we conclude our findings in the last section.

\section{Non-minimally Coupled Scalar Fields in the Jordan Frame revisited}\label{ch2}

Non-minimally coupled scalar fields introduce a unique dynamic to the interaction between matter and gravity. This concept of non-minimal coupling was first introduced into the gravitational action by Pascual Jordan\cite{Jordan-1938,Jordan-1955}. Instead of merely residing within the background spacetime geometry, as described by minimally coupled fields, these scalar fields actively participate in shaping the gravitational field itself. This interaction is governed by a coupling parameter, often denoted as $\xi$, which quantifies the strength of the coupling between the scalar field and spacetime curvature, $R$. In cosmology, non-minimally coupled scalar fields have been instrumental in models of cosmic inflation, which describe the rapid early expansion of the universe. The scalar field responsible for inflation, often called the inflaton, can have a non-minimal coupling to gravity, see Ref.\cite{Bezrukov:2007ep} for inflation driven by the Higgs field. The choice of $\xi$ in such models can significantly impact the inflationary dynamics and the resulting cosmological predictions. In this part, our model is described by the following action:
\ba\label{NMDC action}
S_{\rm NMC}(g)=\int d^4x \sqrt{-g} \bigg[\frac{1}{2}m_{p}^{2}{R} +\frac{1}{2}\xi R\phi^{2}-\frac{1}{2}g^{\mu\nu}\partial_{\mu}\phi\partial_{\nu}\phi-V(\phi) \bigg]\,,
\ea
and we consider a flat FLRW space-time given by
\ba\label{FRW metric}
ds^{2}_{g} = g_{\mu\nu}dx^{\mu}dx^{\nu}=-dt^{2}+a^{2}(t)\delta_{ij} dx^i dx^j\,.
\ea
We then write the original Lagrangian in terms of the point-like parameters characterized by the configuration space, i.e., ${\cal L} = {\cal L}(a, \dot{a}, \phi, \dot{\phi})$. To begin with, we consider the metric (\ref{FRW metric}) and plug into the action (\ref{NMDC action}) to obtain the following point-like Lagrangian, which is suitable for investigation of the symmetry properties of the system:
\ba\label{action NMC}
\mathcal{L}_{\rm NMC}(a,\dot{a},\phi,\dot{\phi})&=&-6a\dot{a}^{2}(1+\xi \phi^{2})-12\xi \phi a^{2}\dot{a}\dot{\phi}+a^{3}\dot{\phi}^{2}-2a^{3}V(\phi).
\ea
Notice that that the number of configuration space (or the minisuperspace) is equal
to two because of the appearance of variables ${a(t), \phi(t)}$ in the system.

\section{Hessian matrix, EL equations \& Non-minimally Coupled Scalar Fields}\label{ch3}

As noticed from Eq.(\ref{action NMC}), the configuration space variables and their time derivative of the models are $q_i=\{a,\phi,\}$ and $\dot{q}^{i}=dq_{i}/dt=\{\dot{a},\dot{\phi}\}$. In the present case, the Hessian matrix can be directly determined to yield
\begin{eqnarray}
[W_{ij}]_{\rm NMC}=\left[
                       \begin{array}{cc}
                         \frac{\partial^{2}{L}}{\partial \dot{a}^{2}} & \frac{\partial^{2}{L}}{\partial \dot{a}\partial \dot{\phi}} \\
                          \frac{\partial^{2}{L}}{\partial \dot{\phi}\partial \dot{a}}  &\frac{\partial^{2}{L}}{\partial \dot{\phi}^{2}}
                       \end{array}
                     \right]=
                    \left[
                       \begin{array}{cc} 
                   -12a(1+\xi\phi^{2})&  -12\xi \phi a^{2} \\  
                    -12\xi \phi a^{2} &  2a^{3} \\
                        \end{array}
                     \right].
\end{eqnarray}
The determinant of the Hessian matrix can be then computed to obtain $-24a^{4}\Big[ 1+\xi\phi^{2}+6\xi^{2} \phi^{2}\Big] \neq 0$ that ensures the crucial property that 
$\mathcal{L}_{\rm NMC}$ is non-degenerate. Thus, the NMC Lagrangian is referred to as regular because the regularity of a Lagrangian is a coordinate-independent property under invertible coordinate transformations \cite{Kurt Symmbook}. It is clearly reduced to the parameters derived in GR case when setting $\xi=0$. Its non-vanishing values typically imply the presence of nontrivial dynamics or gravitational interactions. It's important to emphasize that in gauge field theory, the general solution to the equations of motion includes arbitrary functions of time. Furthermore, the canonical variables are not entirely independent, as they are subject to constraint relations. The energy functional, also known as the Hamiltonian constraint equation, can be derived directly from the canonical momenta through the Legendre transformation, based on the Lagrangian defined by:
\ba\label{energy functional}
\mathcal{H}=E_{\mathcal{L}}=\dot{q}^{i}\frac{\partial \mathcal{L}}{\partial \dot{q}^{i}}-\mathcal{L}.
\ea
In the case of minisuperspace variables in the NMC universe, We find that the Hamitonian function is 
\ba\label{Hamiltonion function NMC}
\mathcal{H}=-6a\dot{a}^{2}(1+\xi\phi^{2})-12\xi \phi a^{2}\dot{a}\dot{\phi}+a^{3}\dot{\phi^{2}}+2a^{3}V(\phi). 
\ea
It is well known that in any cosmic gravitational field theory, $\mathcal{H}=0$  this gives the modified Friedmann equation as,
\ba\label{Modif FM eq}
{H}^{2}&=&\frac{\kappa^{2}_{ \rm eff}}{3}\Big( \frac{{\dot{\phi}}^{2} }{2}+V(\phi)-6\xi\phi\dot{\phi}H\Big), \\
&=& \frac{\kappa^{2}_{ \rm eff}}{3}\Big( \frac{{\dot{\phi}}^{2} }{2}+V(\phi)+\rho_{\rm NMC}\Big),
\ea
where $\rho_{\rm NMC} \equiv -6\xi\phi\dot{\phi}H $  and $\kappa_{\rm eff}=\frac{8\pi G_{\rm N}}{c^{4}(1+\xi \phi^{2})}$ or the effective gravitational constant $G_{\rm eff}=\frac{G_{\rm N}}{1+\xi \phi^{2}}.$ The acceleration equation can be directly obtained from the Eurler-Largrange equation for a(t) and the fluid equation, this gives
\ba\label{acceleration eq.}
\frac{2\ddot{a}}{a}+H^{2}=3H^{2}+2\dot{H}=\frac{1}{(1+\xi \phi^{2})}\Big [ -\frac{1}{2}\dot{\phi}^{2}+V(\phi)-\xi (4H\phi\dot{\phi}+2\dot{\phi^{2}}+2\ddot{\phi}\phi ) \Big].
\ea
It is important to note that the NMC coupling parameter influences the acceleration of the universe's expansion in the following way: when $\xi>0, $  the condition 
$\xi (4H\phi\dot{\phi}+2\dot{\phi^{2}}+2\ddot{\phi}\phi )$  contributes to reducing the rate of acceleration of the universe's expansion. 
Furthermore, we can derive the Euler-Lagrange equation for $\phi$, known as the modified Klein-Gordon equation, as follows:
\ba\label{EL of phi}
\ddot{\phi}+3H\dot{\phi}-\xi R \phi
+V'(\phi)=0.
\ea
With slow-roll condition $\ddot{\phi} \approx 0$ this gives
\ba\label{slow-roll condt}
3H\dot{\phi}-\xi R \phi
+V'(\phi)\simeq 0.
\ea

In the next section, we employ the Noether gauge symmetries to figure out exact solutions of the systems.

\section{Noether gauge symmetries of the NMC action}\label{ch4}
\begin{figure}[!h]	
	\includegraphics[width=9cm]{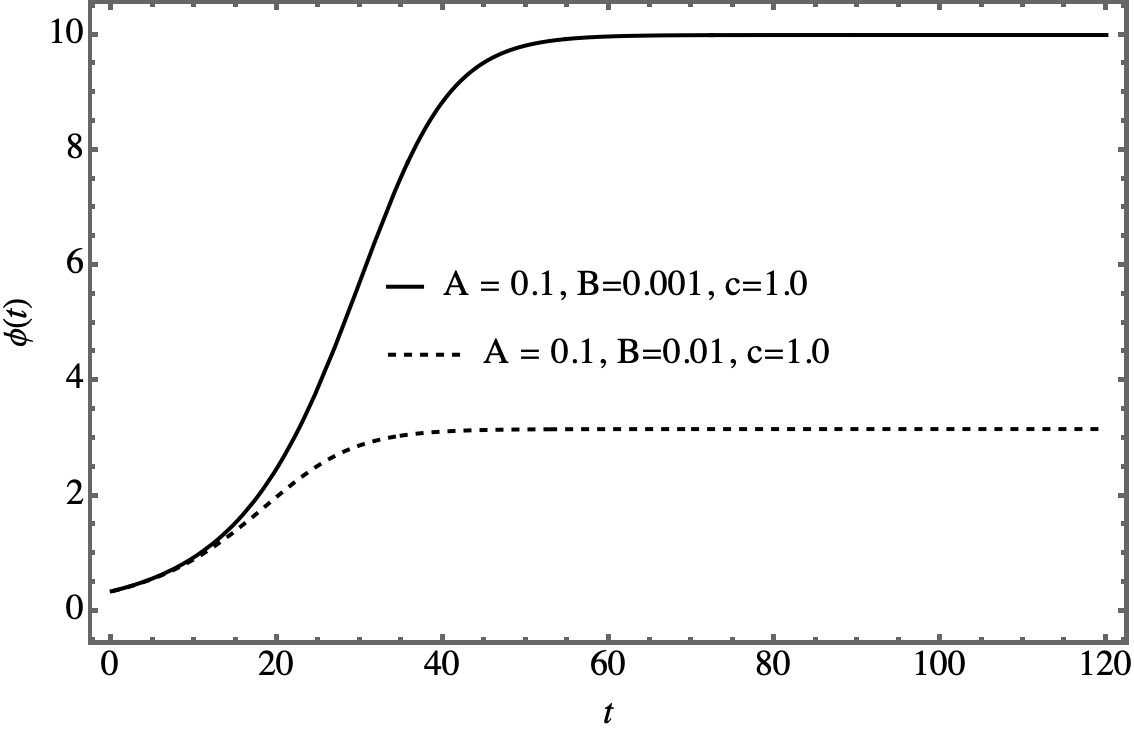}
	\centering
	\caption{The evolution of the solution $\phi(t)$ given in Eq.(\ref{form of scalar field}) for specific values of parameters $A,\,B$ and $c$.}
	\label{phit}
\end{figure}

The Noether gauge symmetry (NGS) technique can be applied to Eq.(\ref{action NMC}) aiming to specify cosmological functions of the NMC gravity. A vector field in this approach can be written as
\begin{equation}\label{vector field NGS}
\mathrm{X}_{\rm NGS}=\tau\frac{\partial}{\partial t}+\alpha\frac{\partial}{\partial a}+\varphi\frac{\partial}{\partial \phi}.
\end{equation}
The first prolongation of $X_{\rm NGS}$ reads
\begin{eqnarray}\label{first prolongation}
 \mathrm{X}^{[1]}_{\rm NGS}&=&\mathrm{X}_{\rm NGS}+\dot{\alpha}\frac{\partial}{\partial \dot{a}}+\dot{\varphi}\frac{\partial}{\partial \dot{\phi}},
\end{eqnarray}
where the undetermined  parameter $\tau$  is a function of  $\{t,a,\phi\}$. The time derivative for  $\alpha(t,a,\phi)$ and $\varphi(t,a,\phi)$ are defined as
\begin{eqnarray}\label{dot alphasa}
\dot{\alpha}(t,a,\phi)&=&\mathrm{D}_{t}\alpha-\dot{a}\mathrm{D}_{t}\tau,\nonumber \\
\dot{\varphi}(t,a,\phi)&=&\mathrm{D}_{t}\varphi-\dot{\phi}\mathrm{D}_{t}\tau.
\end{eqnarray}
Here
$\mathrm{D}_t$ is the operator of a total differentiation with respect to $t$, i.e.
\begin{equation}
    \mathrm{D}_t =\frac{\partial}{\partial t}+\dot{a}\frac{\partial}{\partial a} +\dot{\phi}\frac{\partial}{\partial \phi}.
\end{equation}
The vector field $\mathrm{X}_{\rm NGS}$  is a NGS of a Lagrangian ${\mathcal{L}}(t,a,\phi,\dot{a},\dot{\phi})$, if there exists a boundary term $\mathrm{G}(t, a, \phi)$ \cite{E Noether original } which obeys the Rund-Trautmann identity, see \cite{SARLET-1981,Leone R-2015,Mukherjee:2021bna} for explicit derivation,
\begin{eqnarray}\label{NGS gaugcd}
    \mathrm{X}^{[1]}_{NGS}{\mathcal{L}}+{\mathcal{L}}\mathrm{D}_t\tau=\mathrm{D}_t \mathrm{G}.
\end{eqnarray}
For NSG without gauge term, i.e., $\mathrm{G}=0$, it needs that $\tau=0$. Therefore Eq.(\ref{NGS gaugcd}) is reduced to $\textit{\pounds}_{\mathrm{X}^{[1]}_{\rm NGS}}{\mathcal{L}}=0$ that is the condition for Noether symmetry \cite{Kucuakca:2011np}. The Noether gauge condition yields
\begin{equation}
X^{[1]}_{NGS}\mathcal{L}_{\rm NMC}+\mathcal{L}_{\rm NMC} D_{t}\tau=D_{t}G\,\label{NGS}
\end{equation}
which can be written and distributed for each term in detail as follows:
\ba\label{detail pdes}
\frac{\partial G}{\partial t}+\dot{a}\frac{\partial G}{\partial a}+\dot{\phi}\frac{\partial G}{\partial \phi}&=&\Big(\tau\frac{\partial \mathcal{L}_{\rm NMC}}{\partial t}+\alpha\frac{\partial\mathcal{L}_{\rm NMC} }{\partial a}+\varphi\frac{\partial\mathcal{L}_{\rm NMC} }{\partial \phi}\Big)+\Big(\frac{\partial \alpha}{\partial t} +\dot{a}\frac{\partial \alpha}{\partial a}+\dot{\phi}\frac{\partial \alpha}{\partial \phi} \Big) \frac{\partial \mathcal{L}_{\rm NMC} }{\partial \dot{a}} \nonumber \\ && -\Big( \dot{a}\frac{\partial \tau }{\partial t}+\dot{a}^{2}\frac{\partial \tau}{\partial a}+\dot{a}\dot{\phi}\frac{\partial \tau}{\partial \phi} \Big)\frac{\partial \mathcal{L}_{\rm NMC}  }{\partial \dot{a}}
+\Big( \frac{\partial \varphi}{\partial t}+\dot{a}\frac{\partial \varphi}{\partial a}+\dot{\phi}\frac{\partial \tau}{\partial \phi} \Big)\frac{\partial \mathcal{L}_{\rm NMC} }{\partial \dot{\phi}}  \\ &&
-\Big( \dot{\phi}\frac{\partial \varphi}{\partial t}+\dot{\phi}\dot{a}\frac{\partial \varphi}{\partial  a}+\dot{\phi}^{2}\frac{\partial \varphi}{\partial \phi} \Big)\frac{\partial\mathcal{L}_{\rm NMC} }{\partial \dot{\phi}}+\mathcal{L}_{\rm NMC}\Big( \frac{\partial \tau}{\partial t} +\dot{a}\frac{\partial \tau}{\partial a}+\dot{\phi}\frac{\partial \tau}{\partial \phi}\Big).\nonumber
\ea
We first consider the system given by Eq.(\ref{action NMC}). In this case, we find 38 terms given by
\ba\label{gauge conditon NMC original}
D_{t}G &=& -6a^{2}V(\phi)\alpha-6\alpha\dot{a}^{2}-6\xi\phi^{2}\alpha\dot{a}^{2}-12\xi a\phi\varphi\dot{a}^{2}-2a^{3}\varphi V'(\phi)-24\xi a\alpha \phi\dot{a}\dot{\phi}-12\xi a^{2}\varphi \dot{a}\dot{\phi}+3a^{2}\alpha\dot{\phi}^{2}\nonumber\\&&-12a\dot{a}\alpha_{t}-12\xi\phi^{2}a\dot{a}\alpha_{t} -12\xi a^{2}\phi\dot{\phi}\alpha_{t}-
2a^{3}V(\phi)\tau_{t}+6a\dot{a}^{2}\tau_{t}+6\xi\phi^{2}a\dot{a}^{2}\tau_{t}
+12\xi a^{2}\phi\dot{a}\dot{\phi}\tau_{t}-a^{3}\dot{\phi}^{2}\tau_{t}\nonumber\\&& -12\xi a^{2}\phi\dot{a}\varphi_{t}+2a^{3}\dot{\phi}\varphi_{t}-12a\dot{a}\dot{\phi}\alpha_{\phi}
-12\xi\phi^{2}a\dot{a}\dot{\phi}\alpha_{\phi}-12\xi a^{2}\phi\dot{\phi}^{2}\alpha_{\phi}-2a^{3}V(\phi)\dot{\phi}\tau_{\phi}+6a\dot{a}^{2}\dot{\phi}\tau_{\phi}\nonumber\\&&+6\xi \phi^{2}a\dot{a}^{2}\dot{\phi}\tau_{\phi}+12\xi a^{2}\phi \dot{a}\dot{\phi}^{2}\tau_{\phi}-a^{3}\dot{\phi}^{3}\tau_{\phi}-12\xi a^{2}\phi\dot{a}\dot{\phi}\varphi_{\phi}+2a^{3}\dot{\phi}^{2}\varphi_{\phi}-12a\dot{a}^{2}\alpha_{a}-12\xi\phi^{2}a\dot{a}^{2}\alpha_{a}\nonumber\\&& -12\xi a^{2}\phi\dot{a}\dot{\phi}\alpha_{a}-2a^{3}V(\phi)\dot{a}\tau_{a}+6a\dot{a}^{3}\tau_{a}+6\xi\phi^{2}a\dot{a}^{3}\tau_{a}+12\xi a^{2}\phi\dot{a}^{2}\dot{\phi}\tau_{a}-a^{3}\dot{a}\dot{\phi}^{2}\tau_{a}- 12\xi a^{2}\phi\dot{a}^{2}\varphi_{a}\nonumber\\&& +2a^{3}\dot{a}\dot{\phi}\varphi_{a} .
\ea 
The constraints equations and the PDEs can be expressed as
\ba
\tau_{\phi}=\tau_{a}&=&0,\\
G_{t}+6a^{2}V(\phi)\alpha+2a^{3}\varphi V'(\phi)+2a^{3}V(\phi)\tau_{t}&=&0, \label{pde1} \\
(-\alpha-2a\alpha_{a}+a\tau_{t})(1+\xi \phi^{2})-2\xi a\phi\varphi -2\xi a^{2}\phi\varphi_{a}&=&0,\label{pde2}\\
3\alpha-a\tau_{t}-12\xi \phi\alpha_{\phi}+2a\varphi_{\phi}&=&0,\label{pde3}\\
G_{a}+12a\alpha_{t}(1+\xi\phi^{2})+12\xi a^{2}\phi\varphi_{t}&=&0, \label{pde4} \\
G_{\phi}-2a^{3}\varphi_{t}+12\xi a^{2}\phi\alpha_{t}&=&0, \label{pde5}\\
12\xi \alpha\phi+6\xi a\varphi -6\xi a\phi\tau_{t}+6\alpha_{\phi}(1+\xi\phi^{2})+6\xi a\phi\varphi_{\phi}+6\xi a\phi\alpha_{a}-a^{2}\varphi_{a}&=&0\,. \label{pde6}
\ea
All unknown functions/variables of the above system can be solvable and this just requires a straightforward calculation. 
To solve the above set of equations, let us take a separation of variables and specifically choose for power-law form, i.e. 
\ba
\alpha &=&a^{m}W(\phi),\label{alphaform}\\
\varphi&=& a^{n}Z(\phi).  \label{varphiform}
\ea
and their partial derivative
\ba\label{differential form}
\alpha_{a}&=& ma^{m-1}W(\phi),\quad 
\alpha_{\phi}=a^{m}W'(\phi),\nonumber\\
\varphi_{a}&=& na^{n-1}Z(\phi),\quad\quad
\varphi_{\phi}=a^{n}Z'(\phi).
\ea
Setting $\tau=c_{1}$ and $G=c_{2}$, we then get $\tau_{t}=0$,  $\alpha_{t}=\varphi_{t}=G_{\varphi}=G_{\phi}=G_{t}=0.$ Thus, in this gravity model, the Noether gauge symmetry simplifies to the standard Noether symmetry analysis. For a more comprehensive discussion on this topic, we refer the reader to Ref.\cite{Palia and Tsamparlis2018}, While this outcome is somewhat limiting, we proceed by focusing on the analysis of the cosmological dynamical variables from this point onward.
Starting form Eq.(\ref{pde1}), we then get
\ba\label{sol pde1}
3V(\phi)a^{m}W(\phi)+a^{n+1}Z(\phi)V'(\phi)=0.
\ea
We observe that $m=n+1,$ and then we obtain the following
\ba
V(\phi)=e^{-\int \frac{3W(\phi)}{Z(\phi)}d\phi}.
\ea
Using Eq.(\ref{pde2}), we obtain
\ba
\frac{W(\phi)}{Z(\phi)}=-\frac{2\xi\phi(n+1)}{(1+\xi\phi^{2})(2n+3)}
\ea
Thus, the potential can be written as
\ba
V(\phi)=V_{0}(1+\xi\phi^{2})^{\frac{3(n+1)}{(2n+3)}}.\label{m1}
\ea
It is worth noting that if we set $n=0,$ the potential can be expressed as
\ba
V(\phi)=V_{0}(1+\xi\phi^{2}). 
\ea
An interesting case arises when setting $n=-3$, which gives the scalar potential in the following form: 
\ba\label{n -3 case}
V(\phi)=V_{0}(1+2\xi\phi^{2}+\xi^{2}\phi^{4}).\label{m2}
\ea
This result is quite intriguing. The potential takes a polynomial form similar to the Higgs potential in particle physics, but without a linear term. Additionally, it is symmetric under $\phi \rightarrow -\phi$,
which may have implications for field theory models involving symmetry breaking. In particular, we expanded our analysis around equations (\ref{m1})–(\ref{m2}), where we demonstrate that specific values of $\xi$ yield scalar potentials resembling Higgs-like structures. These forms have significant implications for the shape and steepness of the potential, which in turn affect the duration and dynamics of inflation. We now show more clearly that for small $\xi$, the potential approximates a chaotic (quadratic) form, which supports power-law inflation. For larger values of $\xi$, the potential becomes quartic, influencing both the slow-roll conditions and the predicted scalar spectral index. Multiplying Eq.(\ref{pde3}) by $\xi \phi$ and adding to Eq.(\ref{pde6} ), we then get
\ba
W(\phi)\Big[ -21\xi\phi-\frac{(6\xi-n)(1+\xi\phi^{2})(2n+3)}{2\xi\phi(n+1)} 
+6\xi\phi m\Big]+W'(\phi)\Big[ 36\xi^{2}\phi^{2}+6(1+\xi\phi^{2}) \Big]=0
\ea
Due to $W(\phi)\neq 0$  and $W'(\phi)\neq 0$, hence we get the values of $\xi$ as follows:
\ba
\xi=\frac{4n^{2}-5}{12n+18}
\ea
In the NMC gravity model, we need to assign specific values to $ n \neq -\frac{3}{2} $ and $ n \neq \pm \sqrt{\frac{5}{4}}$ to prevent $\xi$ from taking on zero values. However, it is possible to determine a viable value for $\xi$  from this constraint equation. 
In the case of the inflationary era with the slow-roll condition and $a(t)=e^{H_{0}t}$ where $H_{0}$ is a constant\cite{Harko:2016xip,Darabi:2013caa}, it is very useful to find the evolution of $\phi(t)$ from Eq.(\ref{slow-roll condt}) by substituting $\frac{\dot{a}}{a}={H_{0}}$. Here, we will consider a Higgs-like potential as given in Eq.(\ref{n -3 case}) to determine the form of the scalar field in the context of Noether symmetry. This yields
\ba\label{scalar evolution}
{3H_{0}}\dot{\phi}-{12}\xi\phi H^{2}_{0}+4V_{0}\xi\phi(1+\xi \phi^{2})=0.  \label{ev}
\ea
The evolution equations (particularly Eq. (\ref{ev})) were solved to reveal how $\xi$ directly modifies the scalar field dynamics during inflation.
We also included an analytical solution for $\phi(t)$, which explicitly depends on $\xi$, indicating whether the field rolls slowly or rapidly depending on the sign and magnitude of $\xi$. The above ODE can be analytically solved for $n=1$
to obtain the scalar field during inflationary epoch as
\ba\label{form of scalar field}
\phi(t)=\pm\sqrt{\frac{Ae^{2A(t+c)}}{1+Be^{2A(t+c)}}},
\ea
where $A=\frac{4\xi( 3H^{2}_{0}-V_{0})}{3H_{0}}$ and $B=\frac{4V_{0}\xi^{2}}{3H_{0}}.$ For $A>0,$ when considering only the positive branch of  $\phi(t)$, the scalar field $\phi(t)$ evolves from zero and gradually approaches a finite asymptotic value, i.e. $\phi(\infty)=\sqrt{\frac{A}{B}}=\sqrt{\frac{(3H^{2}_{0}-V_{0})}{V_{0}\xi^{2}}}$ as shown in Fig.\ref{phit}. 
Conversely, allowing both $A$ and $B$ to take negative values by setting $\xi$ and $V_0$ as negative leads to an imaginary value for $\phi(t)$, which is unphysical. Therefore, this case is excluded from our analysis.
For $ n=0$, we get ${3H_{0}}\dot{\phi}-{12}\xi\phi H^{2}_{0}+2V_{0}\xi\phi=0.$  which is in the case of  the chaotic Inflation or the quadratic potential $V(\phi) =V_{0}(1+ \xi \phi^{2}) $ during the inflationary phase where the scalar field exhibits an exponential dynamic behavior with $\phi(t)=\phi_{0}e^{2\xi(2H_{0}-\frac{V_{0}}{3H_{0}})}$. 
Is there a form of the scalar field that depends on the scale factor in a complex manner, derived from an equation involving deeper symmetry principles? Indeed, such a principle exists, known as the Eisenhart lift \cite{Eisenhart origin, Cariglia 2015,K Fin 2018,Finn 2019,Finn 2023 Book,Paliathanasis ESL1,Paliathanasis ELS2,Paliathanasis ELS3,T Chiba 2024 sep,Chiba:2024auh}. Recently, this topic has gained attention in the literature, as evidenced by Ref.\cite{T Chiba 2024 sep,Chiba:2024auh}, which also focuses on the same area of interest. Recent studies have also explored the application of the Eisenhart lift in cosmology across various concepts, as shown in \cite{Carglia2018,Kan2021,Finn2020ic,Paliathanasis2024Eisenhatlinear,Karananas2024,Balcerzak2023,Ben2023}. In this study, we observed that the results obtained for $g(\phi),f(\phi),$ and $V(\phi)$ are in close agreement with those in Ref.\cite{T Chiba 2024 sep,Chiba:2024auh}. However, our work extends this analysis by focusing on a generalized form of the Conformal Killing Equations, incorporating the non-minimal coupling (NMC) constant into the calculations. For further context, Ref.\cite{Paliathanasis ELS3} provides valuable insights into the application of the Eisenhart-Duval lift in cosmological solutions within scalar-tensor theory. This principle plays a key role in the study of equations of motion in both physics and mathematics, expanding equations from lower to higher dimensions. It is particularly useful for analyzing systems with complex symmetries or those involving gravity and the scale factor. In the following section, we will delve deeper into the application of the Eisenhart lift to the Non-Minimally Coupled (NMC) cosmological gravitational field.

\section{Eisenhart Lift in NMC Cosmological Models\label{ch5}}

The application of the Eisenhart liftin gravity enables the study of gravitational interactions within an extended framework. By adding a new coordinate $\chi$ to the NMC gravity model, we can explore the geometrical properties that emerge from the coupling between scalar fields and curvature terms. This approach reveals the interplay between geometry and dynamics, offering deeper insights into gravitational forces and underlying symmetries. The Eisenhart lift method can be use to reformulate the original Lagrangian within this extended framework, as demonstrated below:
\ba\label{Lift L}
\mathcal{L}_{\rm NMC,Lift}=-6a(1+\xi\phi^{2})\dot{a}^{2}-12\xi\phi a^{2}\dot{a}\dot{\phi}+a^{3}\dot{\phi}^{2}+
\frac{\dot{\chi}^{2}}{a^{3}V(\phi)}=\frac{1}{2}G_{AB}\dot{\Phi}^{A}\dot{\Phi}^{A}+\mathcal{W}({\Phi}^{A}),
\ea
where ${\Phi}^{A}=(a,\phi,\chi).$ It can be seen that the field space metric $G_{\rm AB}$ is given by
\begin{eqnarray}
G_{\rm AB}=\left[
                       \begin{array}{ccc} 
                   -12a(1+\xi\phi^{2})&  -12\xi \phi a^{2}& {0}\\  
                  
                   -12\xi \phi a^{2}&  2a^{3} & {0} \\
                  {0} & {0} & \frac{2}{a^{3}V(\phi)}
                        \end{array}
                     \right].\label{Lf1}
\end{eqnarray}
The determinant of this metric is $-\frac{48a}{V(\phi)}(1+\xi\phi^{2}+6\xi^{2}\phi^{2})$ and the inverse metric is
\begin{eqnarray}
G^{AB}=
\left[
\begin{array}{ccc} 
\frac{-1}{12a(1+\xi\phi^2+6\xi^2\phi^2)} & \frac{-\xi \phi}{2a^{2}(1+\xi\phi^2+6\xi^2\phi^2)} & 0 \\  
\frac{-\xi \phi}{2a^{2}(1+\xi\phi^2+6\xi^2\phi^2)} & \frac{1+\xi\phi^{2}}{2a^3(1+\xi\phi^2+6\xi^2\phi^2)} & 0 \\
0 & 0 & \frac{a^3 V(\phi)}{2}
\end{array}
\right].\label{Lf2}
\end{eqnarray}
The Lift Hamiltonian can be derived from 
\ba
\mathcal{H}_{\rm NMC,Lift}&=&\frac{1}{2}G^{AB}p_{A}p_{B},\notag\\
&=& \frac{1}{2}G^{aa}p^{2}_{a}+G^{a\phi}p_{a}p_{\phi}+\frac{1}{2}G^{\phi\phi}p^{2}_{\phi}+\frac{1}{2}G^{\chi\chi}p^{2}_{\chi},\notag\\
&=&\frac{1}{4}\Bigg[-\frac{p^{2}_{a}}{6a(1+\xi\phi^{2}+6\xi^{2}\phi^{2})} -\frac{2\xi \phi p_{a}p_{\phi}}{a^{2}(1+\xi\phi^{2}+6\xi^{2}\phi^{2})}+\frac{(1+\xi\phi^{2})p^{2}_{\phi}}{a^{3}(1+\xi\phi^{2}+6\xi^{2}\phi^{2})}\nonumber\\ &&\quad \quad +a^{3}V(\phi)p^{2}_{\chi}\Bigg]\,,\label{Hamiltonian Lift1}
\ea
where 
\ba
p_{a}&=&\frac{\partial \mathcal{L}}{\partial \dot{a}}=-12a\dot{a}(1+\xi \phi^{2})-12\xi \phi a^{2}\dot{\phi}, \label{pa}\\
p_{\phi}&=&\frac{\partial \mathcal{L}}{\partial \dot{\phi}}=-12\xi\phi a^{2}\dot{a}+2a^{3}\dot{\phi}, \label{p phi} \\
p_{\chi}&=&\frac{\partial \mathcal{L}}{\partial \dot{\chi}}=\frac{2\dot{\chi}}{a^{3}V(\phi)}. \label{p chi2}
\ea
Furthermore, in the Eisenhart-Duval lift formalism, we emphasize how $\xi$ modifies the geodesic equations and contributes to the structure of the field space metric (see Eqs.~(\ref{Lf1})–(\ref{Hamiltonian Lift1})). This geometrical perspective strengthens our understanding of how $\xi$ governs both the inflationary phase and the system’s symmetries. It is revealed by the Hamilton equation that $\dot{p}_{\chi}=-\frac{\partial \mathcal{H}}{\partial \chi}=0$  or $p_{\chi}= \rm {const.}$   By setting  $ p_{\chi}=1$, this gives $\dot{\chi}=\frac{a^{3}}{2}V(\phi).$ 
The non-vanishing Christoffel symbols derived from 
\ba
\Gamma^{C}_{AB}=\frac{1}{2}G^{CD}\Big( \partial_{A} G_{BD}+\partial_{B}G_{AD}-\partial_{D}G_{AB}  \Big),
\ea
are expressed as follows:
\ba\label{CTF}
\Gamma^a_{\ a a} &=& \frac{1}{2a}\frac{(1+\xi\phi^{2}+12\xi^{2}\phi^{2})}{(1+\xi\phi^{2}+6\xi^{2}\phi^{2})}, \\
\Gamma^a_{a \phi} &=& \Gamma^a_{ \phi a} = -\frac{\xi\phi}{2(1+\xi\phi^{2}+6\xi^{2}\phi^{2})},\\ 
\Gamma^a_{\ \phi \phi}& = & \frac{a(1+4\xi)}{4(1+\xi\phi^{2}+6\xi^{2}\phi^{2})},\\
\Gamma^\phi_{ a a} &=&  -\frac{6\xi\phi(1+\xi\phi^{2})}{a^{2}(1+\xi\phi^{2}+6\xi^{2}\phi^{2})},\\
\Gamma^\phi_{ a \phi} &=& \Gamma^\phi_{\ \phi a} = \frac{3(1+\xi\phi^{2})}{2a(1+\xi\phi^{2}+6\xi^{2}\phi^{2})},\\
\Gamma^a_{\ \chi \chi} &= & -\frac{1}{2a^{5}(1+\xi\phi^{2}+6\xi^{2}\phi^{2})V}\Bigg[ \frac{1}{2} +\frac{\xi\phi V'}{V}\Bigg], \\
\Gamma^{\phi}_{\chi\chi} &=& \frac{1}{2a^{6}(1+\xi\phi^{2}+6\xi^{2}\phi^{2})V}\Big[ (1+\xi\phi^{2})\frac{V'}{V} -3\xi\phi\Big],\\
\Gamma^\phi_{\ \phi \phi} & = & \frac{3\xi\phi}{2(1+\xi\phi^{2}+6\xi^{2}\phi^{2})}+\frac{6\xi^{2}\phi}{(1+\xi\phi^{2}+6\xi^{2}\phi^{2})},\\
\Gamma^{\chi}_{a\chi}&=&\Gamma^{\chi}_{\chi a}=-\frac{3}{2a},\, \Gamma^{\chi}_{\phi\chi}=\Gamma^{\chi}_{\chi\phi}=-\frac{V'}{2V}.
\ea
It is worth mentioning that the NMC coupling parameter is a dimensionless quantity; therefore, we can express $1+\xi\phi^{2}+6\xi^{2}\phi^{2}$ accordingly.
 The generalized geodesic equation on the field space manifold can be expressed as
\ba
\ddot{\Phi}^{A}+\Gamma^{A}_{BC}\dot{\Phi}^{B}\dot{\Phi}^{C}=-G^{AB}\partial_{B}\mathcal{W}({\Phi}^{A}).
\ea
However, because of the lift, the NMC Lagrangian contains only kinetic terms and no potential $\mathcal{W}(a,\phi,\chi)$. As a result, we can disregard the term $-G^{AB}\partial_{B}\mathcal{W}$ on the right-hand side. Thus, we can express the geodesic equation in a way that separates each variable as follows:
\ba
\ddot{a}+\Gamma^{a}_{aa}\dot{a}^{2}+\Gamma^{a}_{\phi\phi}\dot{\phi}^{2} +2\Gamma^{a}_{\phi a}\dot{a}\dot{\phi}+\Gamma^{a}_{\chi\chi}\dot{\chi}^{2}&=&0,  \label{geo a}\\ 
\ddot{\phi}+\Gamma^{\phi}_{aa}\dot{a}^{2}+\Gamma^{\phi}_{\phi\phi}\dot{\phi}^{2}+2\Gamma^{\phi}_{a\phi}\dot{a}\dot{\phi} +\Gamma^{\phi}_{\chi\chi}\dot{\chi}^{2}&=&0, \label{geo phi}\\
\ddot{\chi}+2\Gamma^{\chi}_{\phi\chi}\dot{\phi}\dot{\chi}+2\Gamma^{\chi}_{a\chi}\dot{a}\dot{\chi} &=&0.
\ea
\ba
\ddot{a}  + \frac{(1+\xi\phi^{2}+12\xi^{2}\phi^{2})}{2a(1+\xi\phi^{2}+6\xi^{2}\phi^{2})}\dot{a}^{2}+\frac{a(1+4\xi)\dot{\phi}^{2}}{4(1+\xi\phi^{2}+6\xi^{2}\phi^{2})}-\frac{\xi\phi\dot{a}\dot{\phi}}{(1+\xi\phi^{2}+6\xi^{2}\phi^{2})} \nonumber \\ -\frac{1}{2a^{5}(1+\xi\phi^{2}+6\xi^{2}\phi^{2})V}\Big[\frac{1}{2} +\frac{\xi\phi V'}{V} \Big]\dot{\chi}^{2}= 0,\label{Accel eq}\\  
\ddot{\phi} -\frac{6\xi\phi(1+\xi\phi^{2})}{(1+\xi\phi^{2}+6\xi^{2}\phi^{2})}H^{2}+\frac{3\xi\phi\dot{\phi}^{2}}{2(1+\xi\phi^{2}+6\xi^{2}\phi^{2})}+\frac{6\xi^{2}\phi\dot{\phi}^{2}}{(1+\xi\phi^{2}+6\xi^{2}\phi^{2})}+3H\dot{\phi}\frac{(1+\xi\phi^{2})}{(1+\xi\phi^{2}+6\xi^{2}\phi^{2})}\nonumber \\ +\frac{1}{2a^{6}(1+\xi\phi^{2}+6\xi^{2}\phi^{2})V}\Big[(1+\xi\phi^{2})\frac{V'}{V} -3\xi\phi \Big]\dot{\chi}^{2}= 0, \label{Modif KG}\\
\ddot{\chi}  - 3H\dot{\chi} - \dot{\phi} \dot{\chi}\frac{V'}{V}  = 0.
\ea
We can rewrite the last equation in the geodesic equation as
\ba
a^{3}V\frac{\partial}{\partial t}(\frac{\dot{\chi}}{a^{3}V})&=&0,\\
\frac{\dot{\chi}}{a^{3}V}&=& A = \rm const,\\
\dot{\chi}&=&Aa^{3}V(\phi).
\ea
Knowing the functions $a(t), V(\phi)$ and $\phi(t)$ allows us to integrate this equation to find $\chi(t)$.
By substituting $\xi=0$ and using the Eisenhart condition\cite{Finn 2023 Book} $A=\sqrt{2}$ in Eq.(\ref{Accel eq}) and Eq.(\ref{Modif KG}),  we get the acceleration equation  and the Klien-Gordon equation  as in GR case.
\ba
\frac{2\ddot{a}}{a}+H^{2}=-\frac{\dot{\phi}^{2}}{2}+V(\phi),
\nonumber \\
\ddot{\phi}+3H\phi+V'(\phi)=0.
\ea
The next step is to follow the guidelines provided in Ref \cite{Chiba:2024auh,T Chiba 2024 sep} for the conformal Killing equations
\ba\label{Kelling eq} 
\nabla_{A}K_{B}+\nabla_{B} K_{A}=F G_{AB},
\ea
where $\nabla_{A}K_{B}=\partial_{A}K_{B}-\Gamma^{C}_{AB}K_{C}.$ In NMC case we have 6 Killing equations as the following:
\ba\label{Killing eq}
\nabla_{a}K_{a}+\nabla_{a}K_{a}&=&FG_{aa},\\
\nabla_{a}K_{\phi}+\nabla_{\phi}K_{a}&=&FG_{a\phi}=0,\\
\nabla_{a}K_{\chi}+\nabla_{\chi}K_{a}&=&FG_{a\chi}=0,\\
\nabla_{\phi}K_{\phi}+\nabla_{\phi}K_{\phi}&=&FG_{\phi\phi},\\
\nabla_{\phi}K_{\chi}+\nabla_{\chi}K_{\phi}&=&FG_{\phi\chi}=0,\\
\nabla_{\chi}K_{\chi}+\nabla_{\chi}K_{\chi}&=&FG_{\chi\chi}.\nonumber
\ea
This results in a set of Killing equations:
\ba
\partial_{a}K_{a}-\frac{1}{2a}\frac{(1+\xi\phi^{2}+12\xi^{2}\phi^{2})}{(1+\xi\phi^{2}+6\xi^{2}\phi^{2})}K_{a}+\frac{6\xi\phi(1+\xi\phi^{2})}{a^{2}(1+\xi\phi^{2}+6\xi^{2}\phi^{2})}K_{\phi}+6a(1+\xi\phi^{2})F&=&0,\label{eq1 KV}\\
\partial_{a}K_{\phi}+\partial_{\phi}K_{a}+\frac{\xi \phi}{(1+\xi\phi^{2}+6\xi^{2}\phi^{2})}K_{a}-\frac{3(1+\xi\phi^{2})}{a(1+\xi\phi^{2}+6\xi^{2}\phi^{2})}K_{\phi}+12\xi\phi a^{2}F &=&0 \label{eq2 KV},\\
\partial_{\phi}K_{\phi}-\frac{a(1+4\xi)}{4(1+\xi\phi^{2}+6\xi^{2}\phi^{2})}K_{a}-\frac{3\xi \phi}{2(1+\xi\phi^{2}+6\xi^{2}\phi^{2})}K_{\phi}-\frac{6\xi^{2} \phi}{(1+\xi\phi^{2}+6\xi^{2}\phi^{2})}K_{\phi}-a^{3}F&=&0, \label{eq3 KV}\\
\partial_{a} K_{\chi}+\partial_{\chi}K_{a}+\frac{3}{a}K_{\chi}&=&0,\label{eq4 KV} \\
\partial_{\phi}K_{\chi}+\partial_{\chi}K_{\phi}+\frac{V'}{V}K_{\chi}&=&0.\label{eq5 KV},\\
a^{3}V\partial_{\chi}K_{\chi}+\frac{\Big[ \frac{1}{2}+\frac{\xi\phi V'}{V} \Big]}{2a^{2}(1+\xi\phi^{2}+6\xi^{2}\phi^{2})}K_{a}-\frac{\Big[ (1+\xi\phi^{2})\frac{V'}{V}-3\xi\phi \Big]}{2a^{3}(1+\xi\phi^{2}+6\xi^{2}\phi^{2})}K_{\phi}-F&=&0, \label{eq6 KV}
\ea
where $V'=\partial{_\phi}V.$ Here we will allow that 
\ba
K_{a}(a,\phi,\chi),\quad K_{\phi}(a,\phi,\chi),\, \text{\rm and}\quad  F(a,\phi,\chi). \label{gen form of chi}
\ea
By attempting to determine the exponent of $\chi^{\eta}$  in an ansatz form of Eq.(\ref{gen form of chi}) that gives the zero exponent, i.e. $\eta=0$ by the form of the Killing equations itself. This is the  main reason we assume that $K_{a}=a^{\beta}h(\phi),\quad$ $K_{\phi}=a^{\gamma}g(\phi)$ and $F=a^{\tau}f(\phi)$  which can show that
\ba
\partial_{a}K_{a}&=&\beta a^{\beta-1}h(\phi),\quad\quad \partial_{\phi}K_{a}=a^{\beta}\partial_{\phi}h(\phi),\\
\partial_{a}K_{\phi}&=&\gamma a^{\gamma-1}g(\phi),\quad \quad
\partial{_\phi}K_{\phi}=a^{\gamma}\partial_{\phi}g(\phi),\\
\partial_{a}K_{\chi}&=&\frac{-6}{a^{4}V(\phi)},\quad \quad \quad
\partial_{\phi}K_{\chi}=-\frac{2V'}{a^{3}V}.
\ea
We know that $K_{\chi}=\frac{2}{a^{3}V(\phi)}$ due to the fact that $\chi$ does not exist in the metric $G_{AB}$, i.e.
\ba\label{first Killing }
K_{(1)}&=&K^{\chi}_{(1)}\partial_{\chi}=1\partial_{\chi}=\partial_{\chi},\\
K^{\chi}_{(1)}&=&G^{\chi\chi}K_{\chi},\\
1&=&\frac{a^{3}V(\phi)}{2}K_{\chi},
\ea
We can confirm the correctness of the equation solution by substituting $K_{\chi}=\frac{2}{a^{3}V(\phi)}$ into Eq.(\ref{eq4 KV}) and Eq.(\ref{eq5 KV}). 
This gives
\ba
\partial_{a}K_{\chi}+\cancel{\partial_{\chi}K_{a}}+\frac{V'}{V}K_{\chi}&=&-\frac{6}{a^{4}V}+\frac{3}{a}\frac{2}{a^{3}V}=0,\\
\partial_{\phi}K_{\chi}+\cancel{\partial_{\chi}K_{\phi}}+\frac{V'}{V}K_{\chi}&=&-\frac{2}{a^{3}}\frac{V'}{V^{2}}+\frac{2}{a^{3}V}\frac{V'}{V}=0.
\ea
We can verify also that $K^{\chi}p_{\chi}=\frac{2\dot{\chi}}{a^{3}V(\phi)}= \rm constant$ as shown in Eq.(\ref{p chi2}).
From Eq.(\ref{eq1 KV}), this gives
\ba
\beta a^{\beta-1}h(\phi)-\frac{a^{\beta-1}h(\phi)}{2}\frac{(1+\xi\phi^{2}+12\xi^{2}\phi^{2})}{(1+\xi\phi^{2}+6\xi^{2}\phi^{2})}+\frac{6\xi \phi(1+\xi\phi^{2}) a^{\gamma-2}g(\phi)}{(1+\xi\phi^{2}+6\xi^{2}\phi^{2})}+6a^{\tau+1}(1+\xi\phi^{2})f(\phi)=0.\label{rel}  \nonumber \\
\ea
Considering only the exponent of the scale factor term, we get that
\ba
\beta-1=\gamma-2=\tau+1 \label{relation betw}
\ea
which reduces the complexity of Eq.(\ref{eq1 KV})  to
\ba\label{first KV}
f(\phi)=\frac{-\beta h(\phi)+\frac{h(\phi)}{2}\frac{(1+\xi\phi^{2}+12\xi^{2}\phi^{2})}{(1+\xi\phi^{2}+6\xi^{2}\phi^{2})}-\frac{6\xi \phi g(\phi)(1+\xi\phi^{2})}{(1+\xi\phi^{2}+6\xi^{2}\phi^{2})}}{6(1+\xi \phi^{2})}.
\ea
By applying the relation as shown in Eq.(\ref{relation betw}) with Eq.(\ref{eq2 KV}), this gives
\ba\label{second KV}
\gamma g(\phi)+\partial_{\phi}h(\phi)+\frac{\xi \phi h(\phi)}{(1+\xi \phi^{2}+6\xi^{2}\phi^{2})}-\frac{3(1+\xi\phi^{2})g(\phi)}{(1+\xi\phi^{2}+6\xi^{2}\phi^{2})}+12\xi\phi f(\phi)=0. \label{h phi}
\ea
By using Eq.(\ref{relation betw}) to Eq.(\ref{eq3 KV}), this gives
\ba\label{third KV}
\partial_{\phi}g(\phi)-\frac{(1+4\xi)h(\phi)}{4(1+\xi\phi^{2}+6\xi^{2}\phi^{2})}-\frac{3\xi\phi g(\phi)}{2(1+\xi\phi^{2}+6\xi^{2}\phi^{2})}-\frac{6\xi^{2}\phi g(\phi)}{(1+\xi\phi^{2}+6\xi^{2}\phi^{2})}-f(\phi)=0.
\ea
By substituting Eq.(\ref{relation betw}) to Eq.(\ref{eq6 KV}), this allows us to write 
\ba\label{forth KV}
\frac{(1+\frac{2\xi\phi V'}{V})h(\phi)}{4(1+\xi\phi^{2}+6\xi^{2}\phi^{2})}-\frac{\Big[ (1+\xi \phi^{2})\frac{V'}{V}-3\xi\phi\Big] }{2(1+\xi\phi^{2}+6\xi^{2}\phi^{2})}g(\phi)-f(\phi)=0.
\ea
Our approach is to first solve these equations in the GR case by setting $\xi=0$. Hence in GR case Eq.(\ref{first KV}),Eq.(\ref{second KV}),Eq.(\ref{third KV}) and Eq.(\ref{forth KV}) can be written as
\ba\label{set 4 KV GR}
f(\phi)+\frac{\beta h(\phi)}{6}-\frac{h(\phi)}{12}=0.\\
\gamma g(\phi)+\partial_{\phi}h(\phi)-3g(\phi)=0,\\
\partial_{\phi}g(\phi)-\frac{h(\phi)}{4}-f(\phi)=\partial_{\phi}g(\phi)-\frac{h(\phi)}{6}=0,\\
\frac{h(\phi)}{4}-\frac{V'}{2V}g(\phi)-f(\phi)=0.
\ea 
It is straightforward to verify that $\gamma=0,\beta=-1$ results in a trivial solution. To allow $\gamma, \beta,\tau$  to be nonzero and to generalize the three ansatz forms, we make an academic choice of $\gamma=2,\beta=1$ and $\tau=-1$. This results in $\frac{dh(\phi)}{d\phi}=g(\phi)$ and  $f(\phi)=-\frac{h(\phi)}{12}.$ Consequently, we obtain the equation
\ba\label{eq for h phi}
\frac{d^{2}h(\phi)}{d\phi^{2}}-\frac{h(\phi)}{6}=0.
\ea
The  general solution for $h(\phi)$ is
\ba
h(\phi) = C_1 e^{\phi/\sqrt{6}} + C_2 e^{-\phi/\sqrt{6}}.
\ea
Hence $g(\phi)$ can be expressed as
\ba
g(\phi) = \frac{dh(\phi)}{d\phi} = \frac{C_1}{\sqrt{6}} e^{\phi/\sqrt{6}} - \frac{C_2}{\sqrt{6}} e^{-\phi/\sqrt{6}}.
\ea
Whereas the scalar potential in GR case, as predicted by the Eisenhart-Duval lift, is obtained from the equation
\ba
\frac{h(\phi)}{3}=\frac{V'}{2V}g(\phi).
\ea
Hence this gives the relation
\ba
V' = \frac{2V \left( C_1 e^{\phi/\sqrt{6}} + C_2 e^{-\phi/\sqrt{6}} \right)}{3 \left( \frac{C_1}{\sqrt{6}} e^{\phi/\sqrt{6}} - \frac{C_2}{\sqrt{6}} e^{-\phi/\sqrt{6}} \right)}.
\ea
In the view of the Eisenhart-Duval lift the potential is shown in the following form 
\ba
V(\phi) = V_0 e^{-4\phi} \left(e^{\sqrt{6} \phi / 3} - \frac{C_2}{C_1} \right)^{4\sqrt{6}},
\ea
where $V_{0}=e^{-4\sqrt{6}\ln{C_{1}}}.$  In the large field limit during the very early stages of the Universe, assuming $\phi \gg 1$ and $e^{\sqrt{6}\phi/3} \gg \frac{C_{2}}{C_{1}},$ we obtain
\ba
V(\phi) \approx V_{0}e^{-0.73\phi}.
\ea
It is called a runaway potential (also known as Kaluza-Klein-type inflation)  because it has no minimum or stable equilibrium point. As $\phi$ increases, $V(\phi)$ continuously decreases without limit. Such potentials play an important role in cosmology, particularly in relation to the de Sitter swampland conjecture\cite{Ramos 2020}. This conjecture proposes that in a consistent theory of quantum gravity, scalar field potentials should not have stable de Sitter vacua but instead display runaway behavior.\\
For small $\phi$ (weak-field limit), assuming $\phi \ll 1$ ,we expand $e^{\sqrt{6}\phi/3} \approx 1+\frac{\sqrt{6}}{3}\phi$ and define $\lambda=1-\frac{C_{2}}{C_{1}}.$  This gives
\ba\label{Vphi}
V(\phi) \approx V_{0}e^{-4\phi}\lambda^{4/\sqrt{6}}\Bigg(1+\frac{4\sqrt{6}\lambda^{-1}}{3}\phi \Bigg)
\ea
\begin{figure}[!h]	
	\includegraphics[width=9cm]{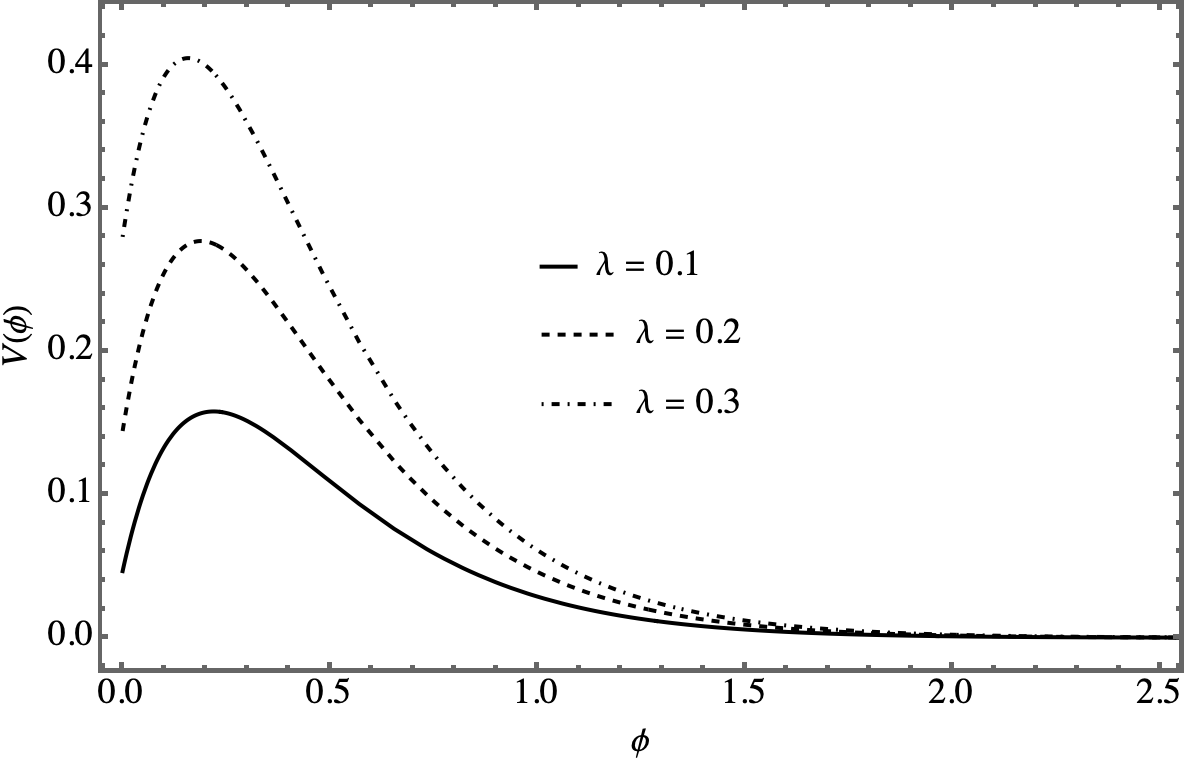}
	\centering
	\caption{The behaviors of the potential $V(\phi)$ as a function of $\phi$ given in Eq.(\ref{Vphi}) using various specific values of a parameter $\lambda$.}
	\label{Vphi1}
\end{figure}
The behaviors of the potential $V(\phi)$ as a function of $\phi$ given in Fig.\ref{Vphi1}. The plot illustrates the potential $V(\phi)$ as a function of the scalar field 
$\phi$ for different values of the coupling parameter $\lambda$. As observed, the potential exhibits a single peak followed by a monotonic decline, characteristic of a hilltop-type potential. Increasing $\lambda$ results in a higher and steeper potential, indicating that the parameter significantly influences the shape and dynamics of the potential. These features may have important implications for the evolution of the scalar field, especially in cosmological scenarios such as inflation, where the steepness and height of the potential affect the slow-roll behavior and duration of inflation. Further expanding the exponential, we obtain the potential in the form:
\ba
V(\phi) \approx V_{0}(1+V_{1}\phi+V_{2}\phi^{2}+...),
\ea
Here, $V_{1},V_{2} ,...V_{n}$ are constant coefficients. This form of the scalar potential is frequently encountered in dark energy models, as it mimics the behavior of a nearly constant potential, enabling it to reproduce effects similar to those of a cosmological constant in the late time universe. In the next section, we will observe that this potential is derived once more using the conformal Killing vector approach, as demonstrated in the following calculations for the weak field limit, as shown in Eq.(\ref{potential Killing}). 

The next part, we will analyze the NMC case under two different limits: the weak field limit, where $1+\xi\phi \ll \xi^{2}\phi^{2}$ by setting $\gamma=2,\beta=1$ and $\tau=-1$. This gives the following  relations:
\ba
f(\phi)=-\frac{h(\phi)}{12}-\xi\phi g(\phi)& = 0 ,\\
\frac{dh(\phi)}{d\phi}-g(\phi)& = 0 , \label{dh phi}\\
\frac{dg(\phi)}{d\phi}-\frac{h(\phi)}{6}-\frac{1}{2}\xi\phi g(\phi)& = 0 , \label{dh phi2}\\
\frac{h(\phi)}{3}-\frac{2V'}{V}\Big[ 1-5\xi\phi\frac{V}{V'}-3\frac{\xi\phi}{2}\frac{V'}{V} \Big] & = 0 . \label{h phi 3}
\ea
Substituting Eq.(\ref{dh phi}) into Eq.(\ref{dh phi2}), this second-order linear variable-coefficient ODE  for $h(\phi)$, i.e. 
\ba
\frac{d^{2}h(\phi)}{d\phi^{2}}-\frac{\xi \phi}{2}\frac{dh}{d\phi }-\frac{h(\phi)}{6} = 0.
\ea
To solve this equation, we employ the power series solution method:
\ba
h(\phi) = \sum_{n=0}^{\infty} a_n \phi^n,
\ea
where $a_{n} $ are coefficients to be determined.
If $\xi$  is small, we can truncate the series after a few terms to derive an approximate solution as follows:
\ba
h(\phi) = a_0 + a_1 \phi + \frac{1}{12} a_0 \phi^2 + \left( \frac{\xi}{2} + \frac{1}{6} \right) \frac{a_1}{6} \phi^3 +\frac{(\xi+\frac{1}{6})a_{0}\phi^{4}}{144} \cdots
\ea
where $a_{0}=h(0)$ and $a_{1}=h'(0)=\frac{dh}{d\phi}$ at $\phi=0.$  
In this approach, we utilize a recurrence relation 
\ba
a_{n+2} = \frac{\frac{\xi}{2} n + \frac{1}{6}}{(n+2)(n+1)} a_n
\ea
to iteratively calculate the coefficients  $a_{n}$.
Substituting $h(\phi)$ into Eq.(\ref{h phi 3}), this yields the scalar potential as shown below
\ba
V(\phi)=V_{0}+V_{1}\phi+V_{2}\phi^{2}+ V_{3}\phi^{3}...\label{potential Killing}
\ea
where $V_{0},V_{1},V_{2},...V_{n} $ are coefficients to be determined,e.g.
\ba
V_{1}&=&\frac{a_{0}V_{0}}{6},\\
V_{2} &=& \frac{V_0}{4} \left( \frac{a_1}{3} + 10\xi + \frac{a_0^2}{18} + \frac{\xi a_0^2}{12} \right).
\ea   
The form of $V(\phi)$ 
could influence the Hubble tension\cite{Hubble tension}, which refers to the discrepancy between the measurements of the Hubble constant from early and late-time observations. Additionally, it could offer a dynamical dark energy model characterized by a time-varying equation of state\cite{DESI}.
The potential satisfies the necessary condition for cosmic acceleration\cite{Faraon book,Faraoni NMC} if $\xi\phi^{2}(1+3\xi)<1$ which gives a constraint on $\xi$ and $\phi$ to ensure that cosmic acceleration occurs. We can determine whether there is a tracking solution for the potential\cite{Steinhardt-1999}, which can be used to avoid the coincidence problem. This is possible if the potential $V(\phi) $ satisfies the required condition.
\ba
\Gamma \equiv V\frac{\frac{d^{2}V}{d\phi^{2}}}{(\frac{dV}{d\phi})^{2}} \geq 1.
\ea
The condition $\Gamma \approx 18 \left( \frac{a_1}{3} + 10\xi + \frac{a_0^2}{18} + \frac{\xi a_0^2}{12} \right) \frac{1}{a_0^2}>1$, this inequality must be satisfied for the tracking solution to exist and avoid the coincidence problem.\\ 
Now we can summarize the form of the (conformal) Killing vectors as follows:
\ba
K_{a,}&=& a^{\beta}h(\phi)= a(t)(a_{0}+a_{1}\phi+\frac{1}{12}a_{0}\phi^{2}),\\
K_{\phi}&=& a^{\gamma}g(\phi)= a^{2}(a_{1}+\frac{1}{6}a_{0}\phi),\\
F(a,\phi)&=& a^{\tau}f(\phi)= a^{-1}\Big[ -a_{0}(\frac{1}{12}+\frac{\xi \phi^{2}}{6}+\frac{\phi^{2}}{144}) +a_{1}(\frac{\phi}{12}-\xi \phi)\Big].
\ea
We can solve the system equation to obtain the Killing vector for the NMC universe as follows:
\ba
K^{a}&= &G^{aa}K_{a}= -\frac{(a_{0}+a_{1}\phi)}{12},\\
K^{\phi}&=&G^{\phi \phi}K_{\phi}= \frac{1}{2a}(a_{1}+\frac{1}{6}a_{0}\phi),\\
K^{\chi}&=& G^{\chi\chi }K_{\chi}=1.
\ea
When expressing a Killing vector field on a coordinate basis, it will be expressed as follows:
\ba
K_{(1)}&=&K^{\chi}\partial_{\chi}=\frac{\partial}{\partial \chi},\\
K_{(2)}&=&K^{a}\partial_{a}+K^{\phi}\partial_{\phi}=
-\frac{(a_{0}+a_{1}\phi)}{12}\frac{\partial}{\partial a}+\frac{1}{2a}(a_{1}+\frac{1}{6}a_{0}\phi)\frac{\partial}{\partial \phi}
\ea
Then we can use these terms to construct the constant of motion:
\ba
\ell_{1}&=&K^{A}_{(1)}p_{A}=K^{\chi}p_{\chi}=p_{\chi}=1,\\
\ell_{2} &=& K^{A}_{(2)}p_{A}= K^{a}p_{a}+K^{\phi}p_{\phi}=-\frac{(a_{0}+a_{1}\phi)}{12}p_{a}+\frac{1}{2a}
(a_{1}+\frac{1}{6}a_{0}\phi)p_{\phi}. \label{ell2}
\ea
Then substituting $ \ell_{2}$ into  the  Lift Hamiltonian for the weak filed as shown in Eq.(\ref{Hamiltonian Lift}) 
\ba
\mathcal{H}_{\rm NMC,Lift}=\frac{1}{4}\Bigg[-\frac{p^{2}_{a}}{6a} -\frac{2\xi \phi p_{a}p_{\phi}}{a^{2}}+\frac{p^{2}_{\phi}}{a^{3}}+a^{3}V(\phi)\Bigg]= 0. \label{Hamiltonian Lift}
\ea
By solving Eq. (\ref{ell2}) and Eq. (\ref{Hamiltonian Lift}) simultaneously, we obtain
\ba\label{pa and p phi}
p_{\phi} &= & C_1 a^{\frac{7}{2}} + C_2 a^{\frac{5}{2}} \phi,\\
p_a &= &-\frac{12}{a_0} \ell_2 + C_3 a^{\frac{5}{2}} + C_4 a^{\frac{3}{2}} \phi = C_3 a^{\frac{5}{2}}  .
\ea
If we assume the term $C_{3}a^{5/2}$ dominates. From Eq.(\ref{pa}) and Eq.(\ref{p phi}), we can rearrange it to yield
\ba\label{devative a phi}
\dot{a} &=& -\frac{12a(1+\xi \phi^2) p_a - 12\xi \phi a^2 \dot{\phi}}{12a(1+\xi \phi^2)},\\
\dot{\phi} &=& \frac{p_{\phi} + 12\xi \phi a^2 \dot{a}}{2a^3}.
\ea
Substituting Eq.(\ref{pa and p phi}) into Eq.(\ref{devative a phi}), hence we obtain
\ba
a(t) &\sim & \left( \frac{24}{C_3} t + \frac{2C}{C_3} \right)^2,\\
\phi(t) &\sim & \frac{C_1}{4} \left( \frac{C_3}{48} t^2 + C_2 t \right) + \phi_0.
\ea
The results given above provide two canonical momenta for the scalar field \( \phi \), the scale factor \( a \), and the Eisenhart lift coordinate \( \chi \), revealing key insights into the conserved quantities $(\ell_{2})$ and dynamics of the non-minimally coupled (NMC) gravity model. 
 
\section{Conclusion}

In this work, we have developed a comprehensive framework for analyzing the dynamics of a non-minimally coupled scalar field (NMC) to gravity by employing the Noether Gauge Symmetry (NGS) method and the Eisenhart-Duval lift. Starting from the point-like Lagrangian, we derived the corresponding Euler–Lagrange equations and constructed the Hamiltonian of the system. The application of the NGS approach led to the identification of exact cosmological solutions and conserved quantities, facilitating the reduction of dynamical equations. One of the key outcomes of this method is the derivation of a generalized power law scalar potential, whose form explicitly depends on the coupling parameter $\xi$. In particular, we showed that for specific values of $\xi$, the scalar potential exhibits a Higgs-like structure, which is symmetric and has potential implications for models of cosmic inflation and symmetry breaking.

The Eisenhart-Duval lift was introduced as a geometric extension of the dynamical system, allowing us to embed the original Lagrangian into an enlarged configuration space. Within this framework, the kinetic sector of the theory is reformulated in terms of a field-space metric, and the geodesic equations governing the dynamics of the system were derived. The Killing vectors associated with this extended geometry yield additional conserved quantities, enriching the symmetry structure of the theory. Importantly, we established a correspondence between the Hamiltonian formulation and the geometric picture provided by the Eisenhart lift, reinforcing the consistency of the approach. The lift condition $A = \sqrt{2}$ was shown to reproduce the canonical structure of general relativity in the minimally coupled limit.

By analyzing the canonical momenta derived from the Killing symmetries, we uncovered an explicit relationship between the scale factor $a(t)$, the scalar field $\phi(t)$, and the coupling constant $\xi$. These relations not only illuminate the role of $\xi$ in driving the inflationary dynamics but also point toward broader applications in early-universe cosmology. Furthermore, we explored the behavior of the system in the weak-field regime, where the scalar potential takes a form compatible with dark energy models and tracking solutions. This suggests that our formalism could be applied beyond the inflationary epoch to investigate late-time cosmic acceleration, potentially offering insights into the Hubble tension and the nature of dynamical dark energy.

In future work, this framework could be extended to include additional fields, anisotropic spacetimes, or higher-order curvature terms. Moreover, the connection between Noether symmetries and swampland criteria offers an intriguing avenue for testing the viability of scalar–tensor theories within the broader context of quantum gravity.

\acknowledgments
N.K. would like to express his sincere gratitude to the anonymous reviewer for pointing out the missing term in the original point-like Lagrangian presented in the first version of this work. This valuable correction allowed the authors to re-examine the study using the original concept in its correct form throughout the entire research. A.T. was supported by Graduate Fellowship (Research Assistant), Faculty of Science, Prince of Songkla University, Contract no.1-2566-02-003. N.K. is funded by National Research Council of Thailand (NRCT) : Contract number N42A660971. The work of P.C. is financially supported by Thailand NSRF via PMU-B under grant number PCB37G6600138.

\end{document}